\documentclass[sigconf]{acmart}

\usepackage{amsthm}
\theoremstyle{theorem}

\theoremstyle{definition}
\newtheorem{definition}{Definition}[section]

\usepackage{multirow}

\usepackage{algorithm}
\usepackage{algorithmicx}
\usepackage{algpseudocode}

\AtBeginDocument{%
  \providecommand\BibTeX{{%
    \normalfont B\kern-0.5em{\scshape i\kern-0.25em b}\kern-0.8em\TeX}}}

\setcopyright{acmcopyright}
\copyrightyear{2020}
\acmYear{2020}
\acmDOI{10.1145/n.m}

\begin{document}

\title{Evolutionary Conflict Checking}

\author{Tao Ji}
\email{taoji@nudt.edu.cn}
\affiliation{%
  \institution{National University of Defense Technology}
  \city{Changsha}
  \state{Hunan, China}
  \postcode{410073}
}

\author{Liqian Chen}
\email{lqchen@nudt.edu.cn}
\affiliation{%
  \institution{National University of Defense Technology}
  \city{Changsha}
  \state{Hunan, China}
  \postcode{410073}
}

\author{Xiaoguang Mao}
\email{xgmao@nudt.edu.cn}
\affiliation{%
  \institution{National University of Defense Technology}
  \city{Changsha}
  \state{Hunan, China}
  \postcode{410073}
}

\author{Xin Yi}
\email{yixin_09@nudt.edu.cn}
\affiliation{%
  \institution{National University of Defense Technology}
  \city{Changsha}
  \state{Hunan, China}
  \postcode{410073}
}

\author{Jiahong Jiang}
\email{jhjiang@nudt.edu.cn}
\affiliation{%
  \institution{Beijing Institute of Tracking and Telecommunication Technology}
  \city{Beijing}
  \state{China}
}








\begin{abstract}
During the software evolution, existing features may be adversely affected by new changes, which is well known as regression errors.
Maintaining a high-quality test suite is helpful to prevent regression errors, whereas it heavily depends on developers.
Continuously augmenting the existing test suite based on the new changes can assist developers in investigating the impact of these new changes. 
And by comparing the executions of the generated test case on two versions, existing techniques are able to detect some common errors.
However, the requirements and oracles on the new changes with existing program behaviors are missing.
In addition, the new changes may introduce new bugs when they are not sufficiently examined with other unchanged code, which finally fails to meet developers' real intentions on changes.
In this paper, we propose the notion of evolutionary conflict checking to validate new changes.
By extracting developers' intention reflected by new changes and transforming the linear evolutionary process into one three-way merge, we detect conflicts between existing behaviors and new changes.
Our experimental results indicate that evolutionary conflict checking is able to be applied for guaranteeing software quality after changes.
\end{abstract}

\ccsdesc[500]{Software and its engineering~Software testing and debugging}
\ccsdesc[500]{Software and its engineering~Software evolution}

\keywords{software evolution, software merging, conflict detection}


\maketitle

\section{Introduction}

Nowadays, developers from different organizations or nations are able to make their own contributions to the same open-source project.
If they are less knowledgeable with the existing changes made by others, regression errors or new bugs are more likely to be introduced once they make new changes.
For example, after adding some new features, developers find that one latent bug exists in the software for a long time and then fix it immediately.
If the developer fails to well examine the relationship between bug fixes and the new features, the new features may be adversely affected (e.g., regression error) or the bug has not been really fixed (e.g., re-opened bug).
For example, Yin et al.~\cite{Yin11} find that, a fundamental reason for developers to commit false bug fixes is that they are not aware of all the potential impacts of the fixes.

Regression testing is widely performed between two versions to prevent regression errors and provides confidence that the new changes does not adversely affect the existing features.
Comparing the execution results of the test suite on two versions, developers determine whether the changed behavior is intended.
In addition, considering that the existing test suite may fail to well cover the changed code, developers need to augment the exiting test suite.
However, it is theoretically infeasible to have a set of test cases that can test new changes properly.
Regression verification is proposed to capture the semantic difference between two program versions, while existing techniques~\cite{Partush13}\cite{Partush14} may fail to output difference precisely. 

Not only regression testing, but also regression verification have the same limitation that oracles between new changes and existing features are missing.
In other words, these techniques cannot determine whether some difference is desired.
For example, during the test suite augmentation, existing methods only can find some common errors such as the null pointer exception~\cite{Xu10}. 
Then, these new generated test cases are provided to developers for further checking, which would increase their burden on maintaining the test suite.
Consequently, the missing of oracles between new changes and existing program behaviors, prevents existing methods to find some more regression errors or new introduced bugs automatically.

\begin{figure}
\centering
\includegraphics[width=1.3in]{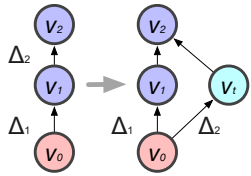}
\caption{Reapplying $\Delta_2$ on $v_0$ to create one 3-way merge.}
\label{transformation}
\end{figure}

Once developers make some changes on the program, they have explicit requirements on the old program behaviors, although they may be less knowledgeable with other parts of the program.
For example, one developer fixes one bug and adds one test case $t$ which can trigger the bug.
This test $t$ should have two different executions $\pi$ and $\pi'$ on the buggy and fixed versions respectively.
And, this abandoned program behavior represented by $\pi$ may exist for a number of program versions.
Till now, it seems that we can leverage existing works~\cite{Li18}\cite{Funaki19} on slicing the commit history to find the dependent commit on which we create one branch to reapply the bug fixes.
However, we still need to make sure that it is proper to recommit latest changes to another earlier version.
As shown in Fig.~\ref{transformation}, there are three program versions $v_0$, $v_1$ and $v_2$.
Intuitively, if we are able to apply the changes $\Delta_2$ on $v_0$ to achieve the same goal as that of applying $\Delta_2$ on $v_1$, and no other undesired behavior is introduced, then we can transform the evolutionary process into one 3-way program merge.
After that, we able to leverage the oracles on merges to detect semantic conflicts between $\Delta_1$ and $\Delta_2$.
And we call these semantic conflicts the \textit{\textbf{evolutionary conflicts}}.

Recently, Sousa et al.~\cite{Sousa18} propose the notion of semantic conflict freedom for 3-way merges, and develop the verification based tool SafeMerge to detect semantic merge conflicts.
Ji et al.~\cite{Ji20} propose test oracles for all merges, and develop TOM to generate tests revealing semantic merge conflicts.
These studies have the potential of being used to detect evolutionary conflicts.

In this paper, we propose the notion of \textit{\textbf{evolutionary conflict checking}} (ECC) to address the problem of finding conflicts between the existing behavior and the newly introduced changes.
By examining the executions of the existing test suite on different program versions, we infer developers' intention on changes.
Then, we propose the contract of determining whether one previous version is the \textit{\textbf{candidate base}} on which we can reapply changes.
Considering the costs of executing tests, we propose the \textit{\textbf{weak evolutionary conflict checking}} to find all of the candidate bases.
In addition, we propose the notion of reordering the commits to support examining the new changes with more previous changes before candidate bases being determined by weak ECC.
This method, which requires more computational resource, is called as the \textit{\textbf{strong evolutionary conflict checking}}.

Based on our algorithms of the weak and strong ECC, we implement jECC to determine the candidate bases for java programs.
In our experiments, we study the evolutionary history of Apache Commons Math.
Experimental results show that more than half (62.2\%) of the commits can be transformed by weak ECC, and strong ECC enables us to examine more previous changes.
After transforming the evolutionary processes into 3-way merges by jECC, we utilize the merge conflict detection tools SafeMerge and TOM to detect the evolutionary conflicts.
And we find one real fixed bug that survived in the studied project for one year.
Considering the high likelihood of reconstructing evolutionary processes, we suggest that developers arrange their development activities in a better way.
Considering that the research on detecting semantic merge conflicts just starts up, we suggest that researchers devote more efforts to improving the effectiveness of detecting merge conflicts.

In summary, our contributions are as follows:
\begin{itemize}
	\item We propose the notion of evolutionary conflict checking to support the software evolution.
	\item We propose the weak ECC to make it practical in real-world development.
	\item We also propose the strong ECC to examine the new changes with more previous changes.
	\item Experimental results show that ECC can help developers find the bugs introduced by new changes.
\end{itemize}

The rest of the paper is structured as follows. 
First we introduce and present out motivations (Section II).
After that, we detail the notion of evolutionary conflict checking (Section III). 
Then we conduct experiments to investigate the effectiveness of our approach (Section IV).
We present the related works (Section V). 
Finally, we conclude (Section VI).

\section{Motivation}
In this section, we briefly present our observation on the software development which motivates our work.

\subsection{Collaborative Development}
Among open-source communities, developers from different national and organizational cultures are involved in the same project, which is well known as the global software development~\cite{global}.

Imagine that one developer is not well familiar with the existing changes made by others, the regression errors or new bugs may be introduced when she/he commits new changes.
To avoid this, for example, the agile software development emphasizes that it is more important to have competent people working together effectively~\cite{agile}.
Although online software development platforms provide the forum to facilitate the communication, developers may have misunderstandings on the intention of existing changes.

It is feasible to maintain a test suite to express the desired program behaviors.
Reruning existing tests is able to prevent some regression errors introduced by new changes, while the test suite is congenitally insufficient.
Hence, when developers rely on existing tests to modify one program version on which they are less knowledgeable, their real intention on new changes might be adversely affected.
In this case, we can tell that this base version is not proper for these new changes.
However, it is impractical for developers to find the proper base version before committing changes.

\subsection{Version Control System}
\begin{figure}
\centering
\includegraphics[width=2.6in]{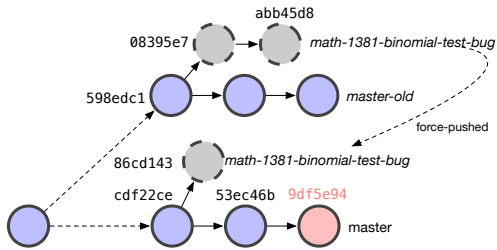}
\caption{The partial commit history of the head and base branches in the pull request \#$43$ of Apache Commons Math.}
\label{rebase}
\end{figure}

Working with version control systems (VCS), developers may pull the remote updates first and then make their changes.
If they work on the previous versions and they want to contribute their changes to the main branch by merging branches, they may have to deal with merge conflicts.
Hence, it makes sense that developers pull the remote updates first and then work on the latest version to avoid conflicts.
On the other hand, the distributed VCSs make it much easier to create branches, which boosts the pull-based development (e.g., Pull Requests of GitHub).

For example, as shown in Fig.~\ref{rebase}, we present the partial commit history of the head and base branches in the pull request \#43 of Apache Commons Math\footnote{https://github.com/apache/commons-math/pull/43}.
The head branch ``math-1381-binomial-test-bug'' is force-pushed after the base branch is force-pushed from ``master-old'' to ``master''.
In general, once core developers decide to accept the changes of the commit ``86cd143''\footnote{https://github.com/apache/commons-math/commit/86cd143. Any commit referred in this paper, can be found via the link ``https://github.com/apache/commons-math/commit/\{SHA\}''. We do not provide links of commits in the remaining paper.}, they would merge the head branch ``math-1381-binomial-test-bug'' into the base branch ``master''.
However, in this case, they closed this pull request and introduced the changes of ``86cd143'' into the master branch, by directly adding one commit ``9df5e94''.
As a result, one 3-way merge is transformed into the linear evolutionary process.
Naturally, we wonder whether the corresponding developers have examined the latest changes of the commit ``53ec46b'' in this main branch carefully.
If not, the latent merge conflicts would degrade the software quality.
Hence, if we want to investigate the conflicts introduced by the commit ``9df5e94'', we need to find the merge base ``cdf22ce'', construct the 3-way merge, and then detect the merge conflicts. 
As we can see, to find the latent conflicts, it is reasonable and necessary to transform the linear evolutionary process back into one 3-way merge.

Based on the above observations, for any new commit, we wonder whether we can find one previous version on which we can reapply the new changes, to achieve the developer's intention (e.g., fixing one bug) while any other undesired program behavior is not introduced.
Then, as shown in Fig.~\ref{transformation}, we transform the evolutionary process to the 3-way merge, then we detect conflicts between new changes and previous changes.

\section{Evolutionary Conflict Checking}
In this section, we introduce the intention on changes, identifying candidate bases and the definition of evolutionary conflicts.
Then, we propose the algorithms of the weak and strong ECC.

\subsection{Intention on Changes}


Working with VCSs, developers briefly describe what they have done in the commit's message, to facilitate code reviewing.
The changed document is not precise enough to help us determine developers' intention on changes.

Not only the source code but also the corresponding test suite is managed by VCS.
Adding new features, fixing bugs and refactoring code are three main development activities.
When developers fix one bug, they normally add one test case into the existing test suite.
And, this test case fails on the buggy version and passes on the fixed version.
When adding new features, developers also add new test cases to describe their requirements on these new features.
Similarly, these newly added test cases fail on the previous version and pass on the new version.
As for code refactoring, relevant test cases should pass on both two versions, while their executions are different in some aspects.
Hence, we consider to infer the developers' intention on changes by examining the executions of test cases.

Test case constructs inputs, invokes the UUTs (Unit Under Testing) and guards the values of output variables.
We execute the test $t$ on program version $v$ and $v'$, and we have two executions $\pi$ and $\pi'$ respectively.
The execution is the process that executes the instructions of a program.
For each terminating execution, we can have the executed sequence of instructions $S$ along with the execution path.
If $S$ or $S'$ covers the code changes $\Delta(v,v')$, we can tell that two executions are non-equivalent due to different instructions included in executions.
As we can see, adding new features, fixing bugs and refactoring code will bring non-equivalent executions of one test case on two versions.
In other words, we may refer developers' intention by identifying those test cases whose executions cover the code changes.
Note that two executions $\pi$ and $\pi'$ can be non-equivalent when they do not cover any changed code, since it is unrealistic to execute one test in the totally same context for several times.
However, this non-equivalence between executions are not brought by code changes.
Hence, we compare executions based on the coverage of changed code, instead of all executed instructions. 
We present the definition of change impacted executions as follows.

\begin{definition}
\label{equivalence}
\textbf{(Impacted Executions)}
Given two program versions $v_i$ and $v_j$, we extract the textual code changes $\Delta(v_i,v_j)$, which consists of a set of deleted lines $L_{src}$ and a set of added lines $L_{dst}$.
We execute one test case $t$ on $v_i$ and $v_j$ to have the terminating executions $t(v_i)$ and $t(v_j)$.
Then, we collect the covered code lines $L_i$ and $L_j$ from executions respectively.
If $L_i \cap L_{src} \neq \emptyset$ or $L_j \cap L_{dst} \neq \emptyset$, we consider that $t(v_i)$ and $t(v_j)$ are impacted by changes, which is denoted as $t(v_i)\neq t(v_j)$.
Otherwise, we consider executions are free of changes' impacts, which is denoted as $t(v_i) = t(v_j)$.
\end{definition}

After one program version $v$ evolves to $v'$, if we have one test $t(v)\neq t(v')$, then we can tell that the program behavior represented by $t(v)$ is abandoned, and the new program behavior $t(v')$ is introduced.
Developers' intention on changes can be represented by the execution pair {<}$t(v),t(v')${>}.

The same test with explicit deterministic inputs may yield different executions on the same program version in the same environment.
These tests are well known as \textbf{\textit{flaky tests}}~\cite{Luo14}, and they have non-deterministic executions for different reasons such as multi-threads, time dependencies, random numbers, etc.
If we fail to collect all of the potential covered lines from one single execution of the flaky test, we may falsely consider $t(v_i)=t(v_j)$.
As a result, to determine whether the changes can be covered, it seems that we have to execute each test for multiple times with expensive costs.

Assumed that one test $t$ is not flaky on $v_i$ but flaky on $v_j$.
In this case, if we have $L_i\cap L_{src} \neq \emptyset$, we have $t(v_i)\neq t(v_j)$.
Otherwise, if we have $L_i\cap L_{src} = \emptyset$, we can tell that $L_j\cap L_{dst} \neq \emptyset$ for any execution of $t$ on $v_j$, since the changed flaky status of $t$ is caused by the code changes $\Delta(v_i,v_j)$.
As a result, if $t$ is flaky on $v_i$ but not on $v_j$, we cannot have $L_i\cap L_{src} = \emptyset$ and $L_j\cap L_{dst} = \emptyset$ at the same time.
In other words, if we have $L_i\cap L_{src} = \emptyset$ and $L_j\cap L_{dst} = \emptyset$, this test must be flaky or not on both versions at the same time.

After executing one flaky test $t$ on $v_i$ and $v_j$ for one single time respectively, if we have $L_i\cap L_{src} = \emptyset$ and $L_j\cap L_{dst} = \emptyset$.
Comparing $L_i$ with $L_j$, if we still have $L_i=L_j$, then we can tell that this test $t$ is not impacted by code changes with more confidence.
For these cases that $L_i\neq L_j$, intuitively we need to spend more computational resource to execute this test for more times.
However, since we have executed it on $v_i$ and $v_j$ respectively and no changed code is covered, we are relatively more confident that this flaky test is not impacted by code changes.
Hence, we think that running one test $t$ on two versions $v$ and $v'$ respectively is able to determine the relationship between $t(v)$ and $t(v')$ with a certain level of confidence.

\subsection{Candidate Bases}

After applying the patch $\Delta(v_n,v_{n+1})$ on $v_n$, $v_n$ evolves to $v_{n+1}$.
As introduced above, each patch consists of the deleted line $L_{src}$ and the corresponding added lines $L_{dst}$.
For each deleted line $l \in L_{src}$, we are able to determine the earliest version that introduces this line into the program by examining the evolutionary history.
After that, we determine a set of versions on each of which we are able to apply the patch $\Delta(v_n,v_{n+1})$.
However, we are not sure that reapplying the patch on different versions is able to achieve the same goals expressed by the evolutionary process from $v_n$ to $v_{n+1}$.

Imagine that developers fix one bug in $v_n$ by fixes $\Delta(v_n,v_{n+1})$.
In most cases, the fixed bug may have existed in several versions.
If we want to reapply the fixes on the previous version, we need to make sure this version has the same bug and the newly created version does not contain this bug now.
As introduced in the above section, we infer developers' intention on changes by examining the executions of one test case on two versions.
Given one test case $t$ and $t(v_n)\neq t(v_{n+1})$, if we want to apply the changes on another version $v_i$ to achieve the same intention, we need to ensure that $t(v_n) = t(v_i)$ and $t(v_{n+1}) = t(v_t)$ (where $v_t$ is the version created by applying changes on $v_i$).

At the same time, after reapplying fixes, we may ask whether the new branch would introduce new behaviors that are not intended.
If one test $t$ does not reveal any difference between $v_n$ and $v_{n+1}$, we think that the execution of $t$ on $v_i$ also should not be affected after applying changes on $v_i$.
Therefore, for one test case $t$ whose executions $t(v_n)=t(v_{n+1})$, it is reasonable that we need to ensure that $t(v_i)=t(v_t)$.
After that, we have a set of versions that may be the candidate bases.
For each candidate, we can create another branch originating from the ancestor and merge these two branches.
We present the formal definition of candidate base as follows.

\begin{definition}
\label{contract}
\textbf{(Candidate Base)}
Suppose that we have one program version $v_n$ and its evolved version $v_{n+1}$. After reapplying textual changes $\Delta(v_n,v_{n+1})$ without textual conflicts on one previous version $v_i$ to have one artificial version  $v_t$, we say $v_i$ is one \textit{candidate base} if the following conditions are satisfied for any test case $t$, where $t(v_n)$ represents the execution of the test case $t$ on the program version $v_n$:

$(t(v_n)\neq t(v_{n+1}) \rightarrow t(v_n)=t(v_i) \wedge t(v_t) = t(v_{n+1}))~~\wedge $

$(t(v_n) = t(v_{n+1}) \rightarrow t(v_i) = t(v_t))$
\end{definition}

\subsection{Evolutionary Conflicts}

If we can find one program version as the candidate base by examining any test theoretically, then we would be able to safely transform the linear evolutionary process into one three-way merge.
At the same time, according to the notion of semantic conflict freedom~\cite{Sousa18} and the test oracles on merges~\cite{Ji20}, we can say that no semantic conflicts exist in this constructed merge.

However, developers rely on the well-written test suite which contains only a limited number of test cases to avoid regression errors, during the real-world software development.
In other words, it makes sense that we only use this limited number of test cases to determine whether one version is the candidate base.
Then, the semantic conflicts may exist in the constructed 3-way merge whose base is determined by the limited number of tests.
We present the definition of evolutionary conflicts as follows.

\begin{definition}
\label{conflicts}
\textbf{(Evolutionary Conflicts)}
Suppose that we have one program version with its test suite ($v_n$, $T_n$) and the evolved version ($v_{n+1}$,$T_{n+1}$).
And we find one previous version ($v_i$,$T_i$) that can be determined as the candidate base by examining any $t\in T_{n+1} \cup T_n \cup T_i$ on the constructed three-way merge.
If any semantic conflicts exist in this merge, we say \textit{evolutionary conflicts} arise after $v_n$ evolves to $v_{n+1}$.
\end{definition}

According to our proposed definition, we are able to identify the evolutionary conflicts in a practical way (i.e., examining the executions of existing test suites on different program versions).

\subsection{Weak Evolutionary Conflict Checking}

After one program version $v_n$ evolves to $v_{n+1}$, the corresponding test suite $T_n$ evolves to $T_{n+1}$ by adding, deleting and modifying test cases.
In real world programs, test cases are methods or functions that construct inputs, invoke other methods and guard values of variables.
By leveraging the static analysis, we are able to identify the newly added, deleted and unchanged test cases.
For example, in Java projects, those unit test cases are stored in the ``test'' directory.
Parsing those test-related files, we are able to extract the dependencies among test classes and methods.
If any dependency of one test is changed, we consider that this test has been changed.
By now, we can classify each test into three categories: added, deleted and unchanged.

Along with the development of open-source communities, continuous integration services that build and test automatically, have been widely applied.
Leveraging the existing process of continuous integration, we can collect tests' executions on their corresponding program versions, to save computational resource.
During the evolutionary process, if one test is not changed, we can have its executions on two versions by examining the results of continuous integration.
As for each of those added and deleted tests, we still need its execution on the other version to investigate the relationship between executions on different versions.

According to our definition of candidate base, we need to evaluate one test's executions on four versions.
However, the dynamic executions of tests are notoriously known as time and resource consuming.
If we already have the coverage information of one test case on one program version, then we can adopt the lightweight technique used by Gligoric et al.~\cite{Gligoric15} to identify those tests that do not need to be re-executed.
If one test case does not cover any changed classes or files, we do not need to rerun it.
Otherwise, we rerun it on the other version.
However, this technique fails to deal with flaky tests.
If we adopt this technique to infer one test's execution on another program version, we are more likely to have the false result $t(v)=t(v')$ since we do not actually run $t$ on $v'$.
Then, we would have a chance of falsely considering one version as the candidate base or missing one real candidate base.
Based on the existence of flaky tests, we can adjust the strategy of inferring one test's execution information.

One project managed by VCS, may have various branches besides one main branch, and each of them originates from some version of another branch.
According to the involvement of other branches in the evolution of one branch, we conclude that committing new changes directly (e.g., manual changes or applying patches) and merging another branch, are two main methods of evolving one branch.
As we can see, each branch consists of an ordered list of program versions.
For example, we can use the Git command ``git log --first-parent'' to return all commits in one branch by order.
To find the candidate bases, we examine the ordered list of versions that appear in the branch of the to-rebase version.

\begin{figure}
\centering
\includegraphics[width=1.8in]{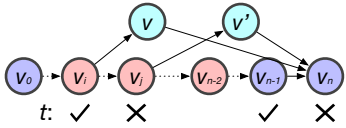}
\caption{Constructing merges on different candidate bases may have different conflicts detected.}
\label{diff_bases}
\end{figure}

Note that, if we have a large number of candidate bases, we need to spend much effort on executing tests on different versions.
Intuitively, can we just choose the earliest candidate base to detect conflicts?
As shown in Fig.~\ref{diff_bases}, there is one test that passes on $v_{n-1}$ and $v_i$, but fails on $v_n$ and $v_j$.
Since this test is not in the written test suite, we can identify all versions from $v_i$ to $v_j$ as the candidate bases.
If we just examine $v_i$ instead of $v_j$, we would miss the evolutionary conflicts revealed by the test $t$.
Considering that reverting changes happens but not often, we suggest that we can start from the earliest version $v_e$ until the given computational resource is consumed.

As shown in Algorithm 1, we present the process of our proposed weak ECC.
After one program version $v_{n-1}$ evolves to $v_n$, the continuous integration service builds the new version.
If the build fails, we skip this version (Lines 3-5).
If not, we continue to analyze the executions of test suites $T_{n-1}$ and $T_n$.
For each $t\in T_{n-1} \cup T_n$, we have its executions on $v_{n-1}$ and $v_n$ (Line 6).
Then, we extract the line-level code changes (comments and other documents are removed) between $v_{n-1}$ and $v_n$ (Lines 7,8).
Then, we identify those tests whose executions $t(v_{n-1})\neq t(v_n)$, and we extract their covered lines from $v_{n-1}$ and $v_n$ respectively (Lines 9-15).
After that, we are able to determine the earliest version $v_e$ that has the code lines identified (Line 16).
Next, we examine each version between $v_e$ and $v_{n-1}$ to construct 3-way merge for checking evolutionary conflicts (Lines 17-30).
For each version $v_i$, we first create another branch based on it (Line 18).
Then we try to reapply the patch $\Delta(v_{n-1}, v_n)$ on $v_i$ (Line 19).
If we fail to reapply the patch, or the patched version fails to be built, we continue to examine the next version (Lines 20-23).
If not, we analyze all of the tests in $T_{n-1}\cup T_n \cup T_i$ and collect the coverage information for them (Line 24).
If any test violates the our contract of candidate bases, we skip this version (Lines 25-28).
Then, we are able to detect evolutionary conflicts for this created 3-way merge.

\begin{algorithm}[h]
\caption{Weak Evolutionary Conflict Checking}
\label{alg:wect}
\begin{algorithmic}[1]

\Function{weak\_ect}{$\mathbb{V}$, $\mathbb{T}$, $n$}
  \State $v_n\mathrm{=}\mathbb{V}[n], T_n\mathrm{=}\mathbb{T}[n],
  ~~v_{n-1}\mathrm{=}\mathbb{V}[n-1], T_{n-1}\mathrm{=}\mathbb{T}[n-1]$
  \If{$!build(\mathbb{V}[n])$}
    \State $exit()$
  \EndIf

  \State $T = update\_coverage(T_{n-1},T_n)$
  \State $patch = \Delta(v_{n-1},v_n)$
  \State $L_{src} = patch.get(``deleted''), L_{dst} = patch.get(``added'')$
  \State $BL_{n-1} = L_{src}, BL_n = L_{dst}$
  \For{ $t \in T$}
    \If{$t.get(v_{n-1})\cap L_{src} \neq \emptyset ~~or~~ 
         t.get(v_n) \cap L_{dst} \neq \emptyset$}
      \State $BL_{n-1} = BL_{n-1} \cup t.get(v_{n-1})$    
      \State $BL_{n} = BL_{n} \cup t.get(v_{n})$    
    \EndIf
  \EndFor
  \State $v_e = blame(BL_{n-1},BL_{n},v_{n-1},v_n)$

  \For{ $v_i \in [v_e, v_{n-2}]$}
  	\State $create\_branch(v_i)$
  	\State $v_t = apply\_patch(v_i,patch)$
    \If{$v_t == null~||~!build(v_t)$}
      \State $delete\_branch()$
      \State $continue$
    \EndIf
    \State $T_{all} = update\_coverage(T, T_i)$
    \If{$violate\_contract(T_{all})$}
      \State $delete\_branch()$
      \State $continue$
    \EndIf
    \State $detect\_conflicts(v_n,v_{n-1},v_i,v_t)$
  \EndFor

\EndFunction

\end{algorithmic}
\end{algorithm}

Note that, executing tests to determine the candidate base is costly.
If we are able to detect the semantic merge conflicts with the lower cost, we can detect conflicts first.
If we find some conflicts, we then determine whether the version is the candidate base.
Otherwise, we continue with the next version.
Hence, we can configure the weak ECC according to the cost and capability of the conflict detection methods.

\subsection{Strong Evolutionary Conflict Checking}

In the process of the weak ECC, we find one earliest version $v_e$ on which reapplying the latest changes $\Delta(v_{n-1},v_n)$ is acceptable.
In other words, we fail to reapply the patch on the parent version $v_{e-1}$ of $v_e$, which means that $\Delta(v_{e-1},v_e)$ leads to the failure of determining $v_{e-1}$ as the candidate base.
We wonder whether there exists some conflicts between the latest changes $\Delta(v_{n-1},v_n)$ and earlier changes $\Delta(v_{i-1},v_i)$ (where $v_i$ is the ancestor of $v_e$).
Before introducing the strong ECC, we present how to reorder commits.

\begin{figure}
\centering
\includegraphics[width=2in]{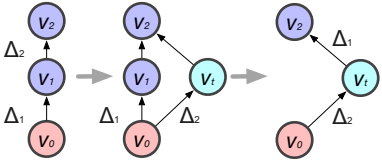}
\caption{If we can transform the evolutionary process into one 3-way merge, then we can reorder the changes.}
\label{ect_extended}
\end{figure}

Imagine that we have three versions $v_0$, $v_1$ and $v_2$, and we can transform this linear evolution process into the three-way merge according to the definition.
Then, as shown in Fig.~\ref{ect_extended}, we tell that applying changes $\Delta(v_1,v_2)$ first is feasible.
Our proposed definition ensures that applying $\Delta(v_1,v_2)$ on $v_0$ is able to achieve the same intention and does not introduce any unintended behavior.
If we are able to reorder these two changes, we must have $t(v_1)=t(v_2)$ when $t(v_0)\neq t(v_1)$.
If this test $t$ satisfies the contract on the candidate base, then we have $t(v_0)=t(v_t)$ and $t(v_t) \neq t(v_2)$, which means that we can apply changes $\Delta(v_0,v_1)$ on $v_t$ to achieve the same intention.
Hence, by using our proposed definition of candidate base, we are able to tell whether changes can be reordered.
If we have determined $v_2$ as the earliest candidate base, then we can tell changes $\Delta_2$ interferes with new changes.
And if we can reorder $\Delta_1$ with $\Delta_2$ and treat the $v_t$ as the new candidate base, then we would be able to examine more previous changes $\Delta_1$ with new changes to detect evolutionary conflicts.

\begin{figure}
\centering
\includegraphics[width=2.8in]{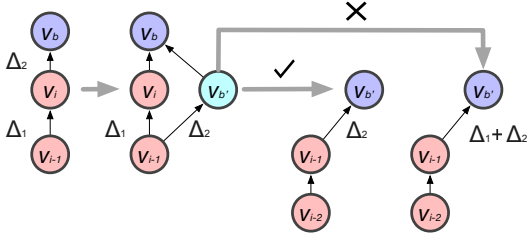}
\caption{Creating one new version as the candidate base.}
\label{sect}
\end{figure}

As shown in the Algorithm~\ref{alg:sect}, we first utilize the weak evolutionary checking algorithm to determine the earliest candidate base version (Line 2).
Then, we examine the earlier version to determine whether we can reorder some changes (Lines 3-28).
Before introducing the details, let us have a look at the Fig.~\ref{sect}.
As shown in Fig.~\ref{sect}, if we can reorder the changes, we can have one new version $v_b'$ which has changes $\Delta_2$ with $v_{i-1}$.
Otherwise, this $v_b'$ should have changes $\Delta_1$ and $\Delta_2$ with $v_{i-1}$.
For the next run, we treat $v_b'$ as the new base version $v_b$ and we can repeat the process.
As shown in Algorithm~\ref{alg:sect}, we check these three versions (Lines 5-16).
If we fail to reorder these changes, we create new $v_b$ by applying changes $\Delta_1$ and then $\Delta_2$, which would omit the version $v_i$ to continue the loop in a easier way.
Otherwise, we treat $v_b'$ as the base version to check whether we can create one 3-way merge to examine evolutionary conflicts (Lines 17-26).
If not, we also apply changes $\Delta_1$ and then $\Delta_2$ on $v_{i-1}$ to have the new $v_b$ (Line 24).
Otherwise, we treat this $v_b'$ as the new base version and then continue the process (Line 27).

Note that, we do not utilize the tests of $T$ (i.e., $T_{n-1}\cup T_n$) to check whether $t \in T$ violates the contracts for the reordered commits.
Once this $t$ violates the contract, we can tell that this $t$ reveals some conflicts in the reordered commits, as is shown in Fig.~\ref{ect_extended}.
According to the test oracles proposed by Ji et al.~\cite{Ji20}, this $t$ describes the unexpected behavior or missing behavior due to merging.
However, as we can see, this $t$ is added and needed by developers.
Hence, we only consider $T$ for checking contracts for the newly created 3-way merge for the changes $\Delta(v_{n-1},v_n)$ (Lines 22-23).


\begin{algorithm}
\caption{Strong Evolutionary Conflict Checking}
\label{alg:sect}
\begin{algorithmic}[1]

\Function{strong\_ect}{$\mathbb{V},\mathbb{T},n$}
  \State $v_e = weak\_ect(\mathbb{V},\mathbb{T},n)$
  \State $v_b = v_e$
  \For{$v_i \in (v_e,v_1]$}
  	\State $create\_branch(v_{i-1})$
    \State $flag=false$
  	\State $v_b' = apply\_patch(v_{i-1},\Delta(v_i,v_b))$

  	\If{$v_b'==null~or~!build(v_b')$}
  	  \State $v_b = apply\_patch(v_{i-1},\Delta(v_{i-1},v_b))$
  	  \State $continue$
  	\EndIf

  	\State $T_{b'} = update\_coverage(T_b,T_i,T_{i-1})$
  	\If{$viloate\_contract(T_b')$}
  	  \State $v_b = apply\_patch(v_{i-1},\Delta(v_{i-1},v_b))$
  	  \State $continue$
  	\EndIf

  	\State $v_{p2} = apply\_patch(v_b',\Delta(v_{n-1},v_n))$
  	\If{$v_{p2}==null~or~!build(v_{p2})$}
  	  \State $v_b = apply\_patch(v_{i-1},\Delta(v_{i-1},v_b))$
  	  \State $continue$
  	\EndIf

  	\State $T_{all} = update\_coverage(T,T_{b'})$
  	\If{$violate\_contract(T_{all})$}
  	  \State $v_b = apply\_patch(v_{i-1},\Delta(v_{i-1},v_b))$
  	  \State $continue$
  	\EndIf

  	\State $v_b = v_b'$, $T_b = T_b'$

  \EndFor

\EndFunction

\end{algorithmic}
\end{algorithm}

If we use the weak ECC continuously, then we are able to reuse the previous results to accelerate the strong ECC to some extent.
Note that the process of reordering commits depends on the contract of the candidate base.
Hence, when we find one version as the earliest candidate base by the weak ECC, then we can reuse the results of finding candidate bases for this version to conduct the strong ECC.
Once we find one other version that cannot be reordered, we cannot reuse previous results and have to examine more with high cost.

\section{Experiments}

In this section, we introduce the implementation of our ECC tool jECC for Java programs.
And then, we study the evolutionary history of the project Apache Commons Math to examine our method's effectiveness by answering three research questions.

\subsection{Implementation of jECC}

We develop jECC for Java programs managed with Git and whose test cases are written by JUnit.
We use jECC to transform the evolutionary history as the first step, and then tools on detecting semantic conflicts for merges can be used to find evolutionary conflicts.

For each program version, we respectively compile the source code and JUnit tests into the \textbf{.jar} archives containing \textbf{.class} files. 
Leveraging the Java class loaders and JUnit, we are able to run the test of one version on different program versions.
Along with the execution of one test, we use JaCoCo\footnote{https://github.com/jacoco/jacoco} to analyze code coverage.
As for the dependencies analysis on tests, we make use of depends\footnote{https://github.com/multilang-depends/depends} to extract dependencies.
Since our algorithm needs to reuse coverage and dependency information of previous versions, we store these relevant information locally to support continuously evolutionary conflict checking.

\subsection{Studied Project}

Apache Commons Math\footnote{https://github.com/apache/commons-math} is a Java mathematical library, and its evolutionary history has been well investigated in different studies~\cite{Just14}\cite{Zhong15}.
Till now, the master branch of this project has a total of 6,495 commits.
We decide to conduct experiments on commits from ``696be68'' to ``68e6de3'', which amounts to 1,791 commits in the first-parent of the ``master'' branch.
And these commits cover the entire phase of the third main version of this studied project.

Math was managed with Subversion (SVN) first and migrated to Git since the commit ``3cbfe27''.
There are a total of 1,660 commits created by using the svn-to-git tool, and 131 commits created by Git.
There are no merges out of 1,660 commits managed with SVN, and 16 merges among 131 commits managed with Git.

\subsection{RQ1: How many commits can be rebased by the weak ECC?}
In this research question, by analyzing the results of the weak ECC, we aim to investigate the potential of creating another development branch based on one prior version rather than working on the latest version.
In our paper, if we can find candidate bases for one commit, we say that this commit is able to be \textit{rebased}.

In our experiments, to reduce the impact of the flaky tests on experimental results, we do not infer one test's new execution by its known executions. 
Hence, by executing tests on all corresponding versions, we identify the candidate bases with a higher level of confidence.

\textit{\textbf{Overview.}}
As shown in Table~\ref{overview}, we present the numbers of commits (Row ``\#C'') that have different numbers of candidate bases (Row ``\#O''). 
The column ``skipped'' includes those commits that fail to be compilable or do not modify any method.
As for the first two commits in the evolutionary history, we do not need to examine them by ECC.
As for the remaining 850 commits, we find that a total of 529 (62.2\%) commits can be rebased according to our proposed definition.
And there are a total of 269 out of 529 (50.9\%) commits that do not have more than five candidate bases.
These results also indicate that the relevant and irrelevant changes are intertwined.

\begin{table}[H]
\caption{The number of commits (Row ``\#C'') that have different numbers of candidate bases (Row ``\#O'').}
\label{overview}
\begin{tabular}{c||r|r|r|r|r|r}
\hline
\#O & skipped & 0    & (0,5] & (5,10] & (10,100] & $>$100 \\ \hline
\#C & 939     & 321  & 269   & 54     & 171      & 35 \\ \hline                     
\end{tabular}
\end{table}

\begin{figure*}
\centering
\includegraphics[width=5in]{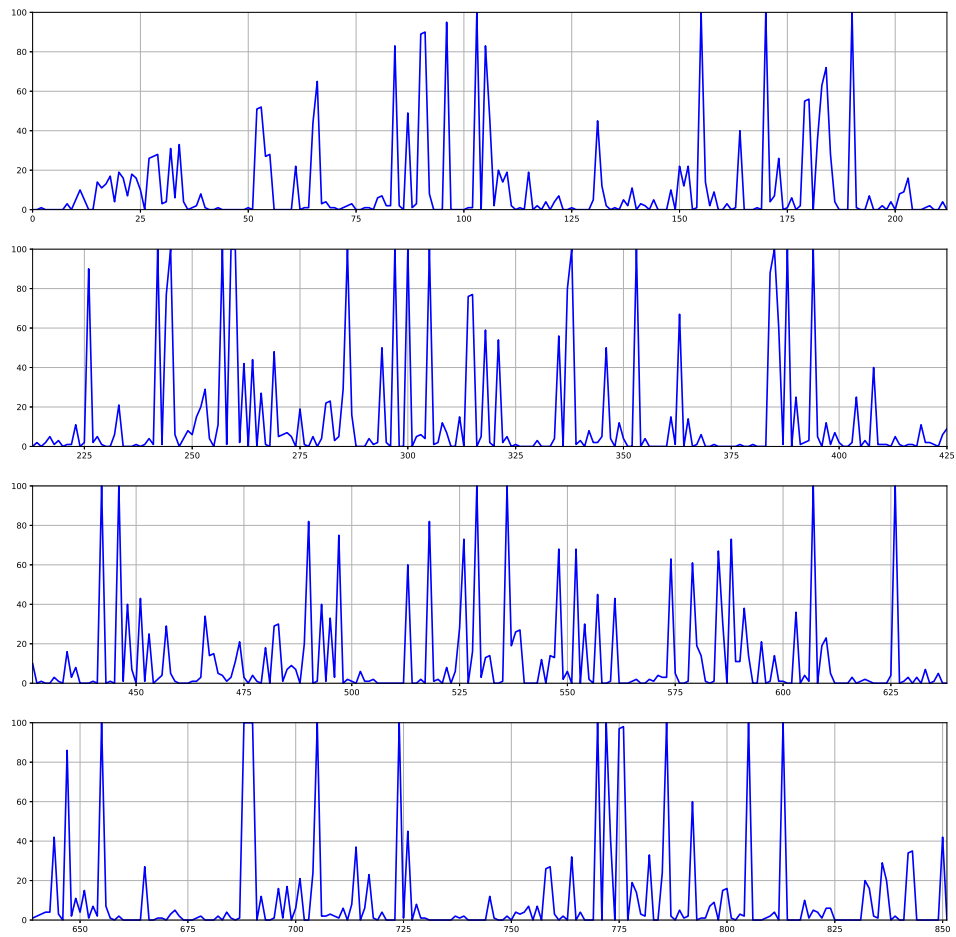}
\caption{The number of candidate bases determined by weak ECC for each commit. X-axis represents the $n$-th commit in order, and Y-axis represents the number of candidate bases.}
\label{weak_nums}
\end{figure*}

We wonder whether those non-rebased commits are gathered together during one short time period to accomplish one development task.
As shown in Fig.~\ref{weak_nums}, we present the number of candidate bases determined by weak ECC for each commit in order.
Manually examining the distribution of the non-rebased commits, we fail to find the strong evidence that developers continuously commit relevant and non-rebased changes individually.
And we find that in different development phases, developers are able to create another branch originating from one previous version for many commits.
As a result, if developers do not utilize the merge, evolutionary conflicts may be prevalent which degrades the software quality.

\textit{\textbf{What commits can be rebased?}} We wonder whether there exists some explicit characteristics in those rebased commits. 
If so, the results would benefit us to arrange the development activities in a much reasonable way.
Intuitively, as introduced above, one developer may continuously commit changes to finish one task in a short time.
In other words, if authors of two adjacent commits are different, these two commits are more likely to have different goals. 
We investigate the rebased commits to examine this assumption.

There are 529 out of 850 (62.2\%) commits that can be rebased by weak ECC according to our definition.
We compare the authors and authored dates between each rebased commit and its parent commit.
There are 273 out of 529 (51.6\%) rebased commits whose author is different from that of the parent commit.
And there are only 79 out of 321 (24.6\%) non-rebased commits whose author is different from that of the parent commit.
Obviously, when the new commit's author is different from the latest commit, this new commit is more likely to be rebased.

Even during the process of completing one development task, the same developers may commit different changes for several times, and these changes may be rebaseable.
For example, five commits whose authors are the same, aim to resolve the issue ``MATH-771'' according to their messages.
The third commit ``610c5e8'' is rebased and the candidate base commit is ``380892d'', which is the first commit for that issue. 

\textit{\textbf{SVN vs. Git}}.
Different from the centralized VCS (e.g., SVN), the distributed VCS (e.g., Git) provides better support for creating and merging branches. 
Intuitively, developers should be much more willing to create another branch originating from the relevant version to complete their development task, instead of continuously committing changes on the latest version.

The results show that, 21 out of 45 (46.7\%) commits directly managed by Git can be rebased.
Comparing to the numbers of rebased commits recreated from SVN, we cannot tell that developers are more willing to find the most proper version and create branches, when they work with Git.
In addition, there are 5 out of 16 merges whose candidate bases are ancestors of the merge bases explicitly determined by Git.
For example, the merge ``92a027b'' have the candidate base ``de73ad3'' while the merge base is ``2fb2221''.

\subsection{RQ2: Does the strong ECC enable us to examine more previous changes?}

In this research question, we aim to investigate the possibility of examining more previous changes.

\begin{figure*}
\centering
\includegraphics[width=5in]{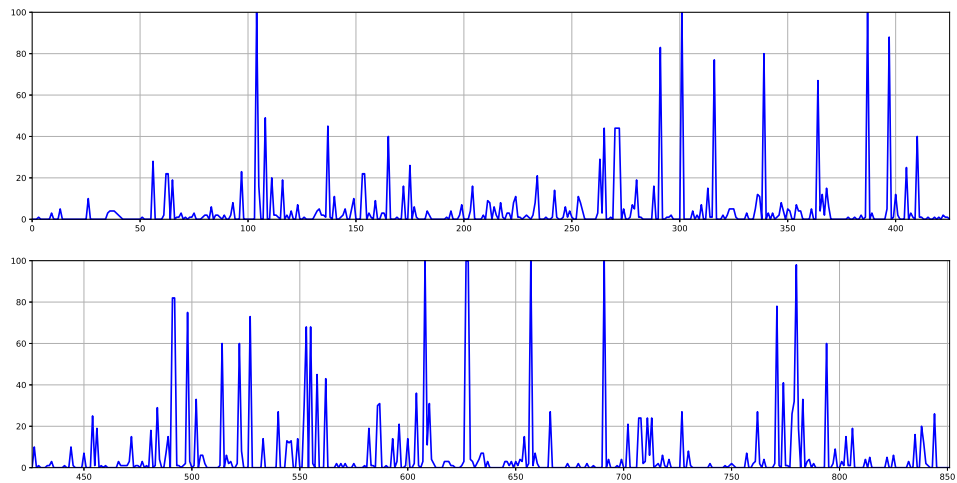}
\caption{The increased number of candidate bases determined by strong ECC for each commit. X-axis represents the $n$-th commit in order, and Y-axis represents the increased number of candidate bases.}
\label{strong_nums}
\end{figure*}

Considering the cost of executing tests, we just use the strong ECC until we find another previous version that cannot be reordered, as is introduced in Section 3.5.
As shown in Fig.~\ref{strong_nums}, we present the numbers of candidate bases added by the strong ECC.
And the strong ECC obviously enables us to examine more previous changes.

\begin{figure*}
\centering
\includegraphics[width=6in]{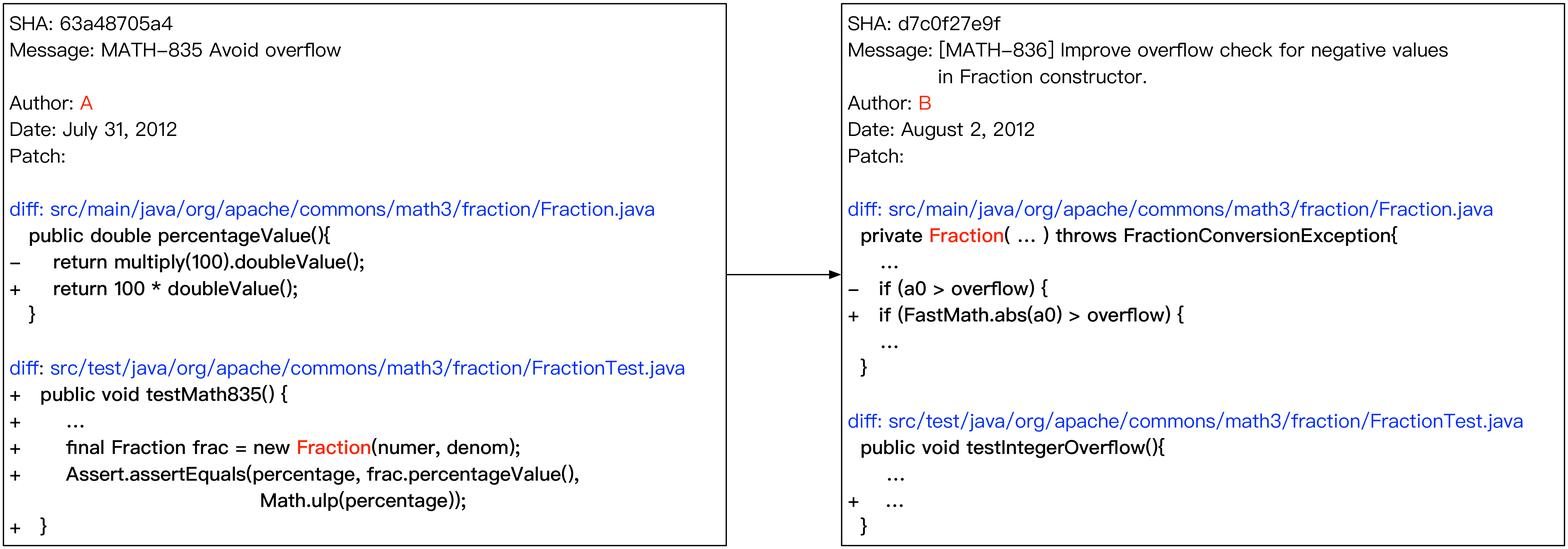}
\caption{The commit ``d7c0f27e9f'' that introduces the evolutionary conflicts.}
\label{real_case}
\end{figure*}

There are a total of 165 commits that fail to be rebased by the weak ECC but can be examined by the strong ECC.
As introduced above in RQ1, supplementary commits may fail to be rebased by the weak ECC.
For example, the commit ``4b42ccd'' modifies the changes introduced by its ancestor commit ``d302ce0a'' which addresses the issue ``MATH-851''.
Note that, the commit ``006f9b7b'' exists between ``4b42ccd'' and ``d302ce0a'', is skipped due to no changes made to source code.  
Our tool jECC examines more changes until the ancestor ``4fd920c'', which also aims to address the issue ``MATH-851''.
The experimental results show that the strong ECC is helpful to examine more previous changes by separating different development goals.

\subsection{RQ3: Are there any evolutionary conflicts detected?}

Recently, Sousa et al.~\cite{Sousa18} develop the verification based tool SafeMerge to verify three-way program merges, and Ji et al.~\cite{Ji20} develop TOM to generate tests to reveal semantic conflicts.
In this research question, we use SafeMerge and TOM respectively to detect the conflicts for created merges.

In our experiments, SafeMerge reports semantic conflict merges created from five commits.
Note that, SafeMerge does not provide any counterexamples revealing the semantic conflicts and false positives may be reported by SafeMerge.
After manually analyzing those conflict merges reported by SafeMerge, without concrete executions triggered by real inputs, we still fail to have enough evidence that evolutionary conflicts actually exist.

Different from SafeMerge, TOM is able to provide generated tests revealing conflicts.
For those merges created by jECC, TOM reports that merges created from two commits hava semantic conflicts.
After manually examining, we find the test generated for revealing conflicts of the merge created from the commit ``7096dabffa'', is flaky due to random numbers generated.
As a result, this generated test fails to ensure the existence of conflicts.
As for the merge created from the other commit ``d7c0f27e9f'' has real conflicts (the conflicts fail to be detected by SafeMerge) revealing by the following generated test:

\noindent\fbox{
\begin{minipage}{3.2in}
public void testPercentageValue()\{\\
$\null\quad$ Fraction f= new Fraction((-1269.30320622502), 0.0, 485);\\
$\null\quad$ Assert.assertEquals((-13644.935128786286),\\
$\null\quad\quad$ f.percentageValue(), 0.01);\\
\}   
\end{minipage}
}

This test case passes on the commit's parent ``63a48705a4'' and fails on previous versions.
This test case reveals one regression error and this regression error exists for one year.
After running this test case on more recent versions, we find this bug has been fixed until the commit ``86545dab3ed''.
The date ranges from August 2, 2012 to September 1, 2013.
And there are 857 commits between these two commits.
As shown in Fig.~\ref{real_case}, the commit ``63a48705a4'' modifies the method ``percentageValue'' which depends on the constructor of this class, as we have to instantiate this class before invoking this method.
Then, another author modifies the constructor of FastMath to address another issue ``MATH-836''.
As the generate test shows, these changes introduce regression errors that fail to be guarded by the added test in commit ``63a48705a4''. 
As we can see, this regression error might have been avoided, if the developer had created one branch based on one early version instead of the current parent commit and conducted semantic conflicts detection.

Considering that detecting semantic conflicts for program merges just starts up, much more work still needs to be devoted to improving existing tools' effectiveness.
For example, SafeMerge fails to detect the conflict detected by TOM, as two branches modify the different methods while SafeMerge does not support this case.
Still, the experimental results show the importance and potentiality of applying ECC on ensuring the software quality.

\subsection{Threats to Validity}

The main threats to the validity of our results belong to the internal and external validity threat categories.

Internal validity threats correspond to the implementation of jECC and the relevant scripts. Although we have reviewed the implementation carefully, the bugs may exist and threat to the validity of results.

External validity threats correspond to the project analyzed.
Projects developed with other software engineering methods, may have less or more evolutionary conflicts.
Our proposed evolutionary conflict checking still can be used to ensure the software quality.

\section{Related Work}
In this section, we present related work on regressions, incorrect bug fixes, and merge conflicts.

\subsection{Regressions and Incorrect Fixes}

Zimmermann et al.~\cite{Zimmermann12} study reopened bug reports of Microsoft Windows, and conclude the main reasons that bugs get reopened as follows: (1) poor or incorrect bug fixes; (2) regressions (i.e., the fixed bug reappears after more changes are made); and (3) the general process of fixing bugs. 
Yin et al.~\cite{Yin11} conduct a study on incorrect fixes from both commercial and open-source projects.
Their study results show that at least 14.8\%\textasciitilde 24.4\% examined fixes are incorrect, and developers who commit and review fixes are not well familiar with relevant code.

Validating new changes properly and sufficiently, is the main goal of guaranteeing the quality of evolving software.
Along this direction, regression testing and regression verification are proposed.
Most of the existing work on regression testing focus on reducing the cost and expanding the test suite.
Yoo and Harman~\cite{Yoo12} conduct an survey on regression testing minimization, selection and prioritization, in which relevant advanced techniques are introduced.
Zhu et al.~\cite{Zhu2019} propose one framework for checking the regression test selection tools.
Regression test generation attempts to generate new tests exposing the behavioral differences between both versions. 
Taneja and Xie~\cite{Taneja08} construct the conditional statement which compares the return values of both versions of the target method, and then generate tests that cover the different branches in the driver method.  
Person et al.~\cite{Person11} propose directed incremental symbolic execution to find those path conditions affected by code changes. 
Xu et al.~\cite{Xu10} identifies code and existing test cases that are affected by changes, then these identified tests are used to seed the concolic or genetic test case generation approach to create new test cases.

As Jin et al.~\cite{Jin10} explain, the generated test cases may not reveal the regression faults while cover the changed parts, due to the missing oracles on changed parts.
Hence, they develop BERT to generate test inputs that cover different parts and analyze the behavioral differences.
After that, with the participation of developers, the regression faults are able to be identified.

Regression verification, which is also well known as proving program equivalence, attempts to leverage formal verifications.
Partush and Yahav~\cite{Partush13} construct the correlating program from two program versions, and use a correlating abstract domain to compute the relationships between variables.
Felsing et al.~\cite{Felsing14} propose to reduce the equivalence of two related programs to Horn constraints and use SMT solvers to solve the constraints.
Churchill et al.~\cite{Churchill19} propose to construct semantics-driven product programs by using alignment predicates and trace alignments, and then use tests to learn the invariants that would be inputted into an SMT solver.

Different from those works that focus on exploring the difference between versions, other works on identifying incorrect bug fixes start from some basic requirements on fixes.
Gu et al.~\cite{Gu10} define two criteria coverage and disruption to determine whether a fix resolves a bug.
The coverage measures the extent to which the fix correctly handles all bug-triggering inputs, and the disruption counts the regressions by rerunning the test suite.
Le and Pattison~\cite{Le14} propose the multiversion interprocedural control flow graph, and apply static analysis on this representation to detect common errors introduced by patches, such as integer overflows, buffer overflows and null-pointer deferences.

In our paper, different from those introduced related works, we investigate the evolutionary history of programs, and propose the notion of transforming the linear evolutionary process into the 3-way merge, by focusing on the relationship between existing behaviors and new changes.
And our proposed approach is able to be applied to solve the problems of finding regressions and incorrect changes that are introduced due to conflicts between existing program behaviors and new changes.

\subsection{Merge Conflicts Detection}
Mens~\cite{mens02} provides a comprehensive summary of excellent merging techniques such as textual, syntactic, semantic, structural and operation-based merging.

After decades, developers still rely on textual merge tools to deal with daily work on merges.
And Ahmed et al.~\cite{Ahmed17} study the relationship between code smells and merge conflicts, and results show merges contain more code smells when conflicts arise.
Mckee et al.~\cite{McKee17} conduct interviews of 10 software practitioners to understand their perspectives on merge conflicts and resolutions. 
According to the unmet needs of software practitioners, they suggest researchers and tool builders focus on program comprehension, history exploration, etc. 
Nishimura and Maruyama~\cite{Nishimura16} present one tool that exploits the fine-grained edit history to assist developers to examine the merge conflicts.

For the last decade, some new ideas and trends have emerged.
Semi-structural merging~\cite{Apel11} aims to achieve the balance between generality and performance.
Proactive or early detection of conflicts~\cite{Brun11}\cite{Guimaraes12}\cite{Nguyen15} is proposed to decrease the possibility of conflicts when developers really merge branches.
To assist developers in resolving conflicts, interactive approaches that ranks the conflicts resolutions have been proposed~\cite{Nan12}\cite{Zhu2018}.
Xing and Maruyama~\cite{Xing19} introduce the automatic program repair techniques to resolve the merge conflicts by leveraging the existing test cases.

Based on earlier works~\cite{Horwitz89}\cite{Yang90}, Sousa et al.~\cite{Sousa18} propose the contract of semantic conflict freedom, and then propose the verification on three-way merges to increase developers' confidence on the merge result with respect to the contract. 
Ji et al.~\cite{Ji20} propose the test oracles for 2-way, 3-way and octopus merges, and they develop TOM which is built on top of EvoSuite to generate tests revealing semantic merge conflicts.
In this paper, we utilize the most recent tools SafeMerge and TOM to detect evolutionary conflicts after transforming the linear evolutionary process into one 3-way merge.

\section{Conclusion}
In this paper, we propose the notion of evolutionary conflict checking, to address the problem of validating new changes.
And considering the costs of determining candidate bases, we propose the weak ECC and strong ECC respectively.
As for reconstructing the evolutionary processes, we develop jECC for java programs managed with Git.
In the experiments, we study the evolutionary history of Apache Commons Math, and the experimental results show that more than half of the commits can be rebased to conduct evolutionary conflict checking.
By using the merge conflict detection tool TOM, we find one real case that has evolutionary conflicts.
Experimental results suggest that we need to apply ECC in real-world development and devote much more efforts to improving the merge conflict detection.

\bibliographystyle{ACM-Reference-Format}
\bibliography{references}


\end{document}